\documentstyle[12pt]{article}
\begin{document}
\def\beq{\begin{equation}}
\def\eeq{\end{equation}}
\def\bea{\begin{eqnarray}}
\def\eea{\end{eqnarray}}
\def\ve{\vert}
\def\nnb{\nonumber}
\def\ga{\left(}
\def\dr{\right)}
\def\aga{\left\{}
\def\adr{\right\}}
\def\rar{\rightarrow}
\def\nnb{\nonumber}
\def\la{\langle}
\def\ra{\rangle}
\def\ba{\begin{array}}
\def\ea{\end{array}}

\title{ {\small { \bf $B_q \rar l^+ l^- \gamma$ DECAYS
                      IN LIGHT CONE QCD} } }

\author{ {\small T. M. ALIEV \thanks
{e-mail:taliev@rorqual.cc.metu.edu.tr}\,\,,
A. \"{O}ZP\.{I}NEC\.{I} \thanks 
{e-mail:e100690@orca.cc.metu.edu.tr}\,\, and \,\,
M. SAVCI \thanks {e-mail:savci@rorqual.cc.metu.edu.tr}} \\
{\small Physics Department, Middle East Technical University} \\
{\small 06531 Ankara, Turkey} }

\begin{titlepage}
\maketitle
\thispagestyle{empty}

\begin{abstract}
\baselineskip  0.7cm
The radiative dileptonic decays $B_s(B_d) \rar  l^+ l^- \gamma ~
(l = e,~\mu)$ are investigated within the Standard Model . The transition
formfactors are calculated in the framework of the light cone QCD sum
rules method and it is found that the branching ratios are 
$B(B_s \rar e^+ e^- \gamma) =2.35 \times 10^{-9}$, $B(B_s \rar \mu^+ \mu^- \gamma)
=1.9 \times 10^{-9}$, $B(B_d \rar e^+ e^- \gamma) = 1.5 \times 10^{-10}
$ and $B(B_d \rar \mu^+ \mu^-
\gamma) = 1.2 \times 10^{-10}$. A comparison of our results with
the constituent quark model predictions on the branching ratios is
presented.

\end{abstract}
\vspace{1cm}
~~~~~~~PACS numbers: 12.38.Bx, 13.20.He, 13.25.Hw
\end{titlepage}
\baselineskip  .7cm
\newpage

\setcounter{page}{1}
\section{Introduction}
The Flavour Changing Neutral Current (FCNC) processes are  the
most promising field for testing the Standard Model (SM) predictions
at loop level and for establishing new physics beyond that (for a   
review see \cite{R1} and references therein). At the same time the  
rare decays
provide a direct and reliable tool for extracting information
about the fundamental parameters of the Standard Model (SM), such as
the Cabibbo-Kobayashi-Maskawa (CKM) matrix elements $V_{td},~V_{ts}$
and $V_{ub}$ \cite{R2}.

After the experimental observation of the $b \rar s \gamma$
\cite{R3} and $B \rar X_s \gamma$ \cite{R4} processes, the interest
is focused on the other possible rare $B$-meson decays, which are  
expected to be observed at future $B$-meson factories and fixed target
machines. In addition to being used in the determination of the CKM matrix
elements, the rare $B$-meson decays could play an important role
in extracting information about some hadronic parameters, such as
the leptonic decay constants $f_{B_s}$ and $f_{B_d}$.
Pure leptonic decays
$B_s \rar \mu^+ \mu^-$ and $B_s \rar e^+ e^-$ are not
useful for this purpose, since these decays are helicity suppressed and
as a result they have branching ratios $B(B_s \rar  \mu^+ \mu^-) 
\simeq 1.8 \times 10^{-9}$ and $B(B_s \rar  e^+ e^-) \simeq 
4.2 \times 10^{-14}$ \cite{R5}. For $B_d$ meson case the situation
becomes worse due to the smaller CKM angle. Although the process  
$B_s \rar \tau^+ \tau^-$, whose branching ratio in the SM is      
$B(B_s \rar \tau^+ \tau^-) = 8 \times 10^{-7}$ \cite{R6}, is free of
helicity suppression, its observability is expected to be compatible with
the observability of the $B_s \rar \mu^+ \mu^-$ decay only 
when its efficiency is better than $10^{-2}$.   

When a photon is emitted in addition to the lepton pair, no helicity
suppression exists anymore and larger branching ratios are expected.
For that
reason, the investigation of the $B_{s(d)} \rar l^+ l^- \gamma$ decay
becomes interesting. 
The branching ratios of these processes depend quadratically on the leptonic
decay constants
of B mesons and hence it could be  a possible alternate in determining
$f_{B_s}$ and $f_{B_d}$.
In \cite{R7}, these decays are investigated in the SM using the
constituent quark approach and it is shown that the diagrams
with a photon radiation from the light quark give  the dominant
contribution to the decay amplitude which is inversely proportional
to the constituent light quark mass. However  the concept of the   
"constituent quark mass" is itself poorly understood. 
Therefore, any prediction on the branching ratios,
in the framework of the above mentioned approach,
is strongly model dependent. 

In this work, we investigate the $B_{s(d)} \rar l^+ l^- \gamma$
processes 
practically in a model independent way, namely, within the framework
of the light cone QCD sum rules method (more about the method and its
applications can be
found in a recent review \cite{R8}). The paper is organized as
follows: In sect.2 we give the relevant effective Hamiltonian for the
$b \rar q l^+ l^-$ decay. In sect.3 we derive the sum rules
for the transition formfactors. Sect.4 is devoted to the numerical
analysis of  the formfactors, and the calculation of the differential 
and total widths for the $B_q \rar l^+ l^- \gamma$ ($q=s,~d$) decays.
In this section
we also present a comparison of our results 
with those of \cite{R7}.

\section{Effective Hamiltonian}

The most important contribution to $B_q \rar l^+ l^- \gamma$ 
$( l = e ,\mu)$ stems from the effective Hamiltonian which
induces the pure leptonic process $B_q \rar \l^+ \l^-$ .  The short distance
contributions to $b \rar  l^+ l^- q$ decay, comes from  the box,
Z-boson and photon mediated diagrams (Fig.1).  The QCD corrected quark level
amplitude in the SM can be written as \cite{R9,R10} :

\bea
{\cal M} &=& \frac{\alpha G_F}{\sqrt{2} \pi} V_{tb} V_{tq}^*
\Bigg{[} C_9^{eff} (\bar q \gamma_\mu P_L b) \bar l \gamma_\mu l +
C_{10}\bar q \gamma_\mu P_L b \bar l \gamma_\mu \gamma_5 l - \nnb \\
&& - 2 \frac{C_7}{p^2} \bar q i \sigma_{\mu \nu} p_\nu ( m_b P_R + m_q P_L ) b
\bar l \gamma_\mu l ~.
\Bigg{]}
\eea

Here $P_{L(R)} =\left[ 1 - (+) \gamma_5 \right] /2$ , and p is the momentum
of the lepton pair. The analytic expressions for all Wilson coefficients can
be found in \cite{R9,R10} .  In further considerations we shall neglect the mass
of the light quarks.

As we have already noted, the pure leptonic processes $B_q \rar l^+ l^-$  $(l=e, \mu)$ are
helicity suppressed.  If a photon is attached to any of the charged lines in
Fig.1, the situation becomes different; helicity suppression is overcome.
If a photon is emitted from the final charged lepton lines, it follows from the helicity
arguments that the amplitude of such diagrams must be proportional to the lepton mass 
$m_l~( l = e ,\mu)$. Therefore the contribution of such diagrams are negligible. 
When a photon is attached to any charged  internal line, the contributions of these
diagrams will be strongly suppressed by a factor of $m_b^2/m_W^2$
in the Wilson coefficients, since the resulting operators have dimension 8,
which are two orders higher than usual operators in (1).
So, we conclude that the main contribution comes from the
diagrams in Fig.1 with a photon radiation from the initial quark lines.
Thus the corresponding matrix element
for the process $B_{s(d)} \rar  l^+ l^- \gamma$ can be written as
\newpage
\bea
\la \gamma \ve {\cal M} \ve B \ra &=&
\frac{\alpha G_F}{2 \sqrt{2} \pi} V_{tb} V_{tq}^* \Bigg{\{} C_9^{eff} \bar l
\gamma_\mu l \la \gamma(q) \ve \bar q \gamma_\mu (1-\gamma_5) b \ve B(p+q) \ra
+ \nnb \\ 
&& + C_{10}^{eff} \bar l \gamma_\mu \gamma_5 l \la \gamma(q) \ve \bar q
\gamma_\mu(1-\gamma_5) b \ve B(p+q)\ra - \nnb \\ 
&& - 2 C_7 \frac{m_b}{p^2} \bar l
\gamma_\mu l \la \gamma(q) \ve \bar q i \sigma_{\mu \alpha} p_{\alpha} (1+
\gamma_5) b \ve B(p+q) \ra \Bigg{\}}  
\eea

These transition amplitudes can be written in terms of two independent, 
gauge invariant (with respect to the electromagnetic field) structures:

\bea
\la \gamma(q) \ve \bar q \gamma_\mu ( 1- \gamma_5) b \ve B(p+q) \ra &=& 
e \Bigg{\{}
\epsilon_{\mu \alpha \beta \sigma} e_\alpha^* p_\beta q_\sigma
\frac{g(p^2)}{m_B^2} +\nnb \\
&& + i \left[ e_\mu^* (p q ) - (e^* p) q_\mu\right] \frac{f(p^2)}{m_B^2}
\Bigg{\}}~, \nnb \\
\la \gamma(q) \ve \bar q i \sigma_{\mu \alpha} p_\alpha (1+ \gamma_5) b \ve
B(p+q) \ra &=& e \Bigg{\{} \epsilon_{\mu \alpha \beta \sigma} e_\alpha^*
p_\beta q_\sigma \frac{g_1(p^2)}{m_B^2} +\nnb \\
&& + i \left[ e_\mu^* (p q) - (e^* p ) q_\mu \right] \frac{f_1(p^2)}{m_B^2} \Bigg{\}}~.
\eea
Here, $e_\mu$ and $q_\mu$ stand for the polarization vector and the momentum
of the photon, $p$ is the momentum of the lepton pair, $g(p^2)$ ,
$g_1(p^2)$ , and $f(p^2)$ , $f_1(p^2)$ describe the parity conserving and
parity violating formfactors.  Thus, the main problem is to calculate the
formfactors $g$ , $g_1$ and $f$ , $f_1$ including their momentum dependence.
For this aim we will employ the light cone QCD sum rules method.

Note that the formfactors $g$ and $f$ are calculated in the light cone QCD 
sum rules in \cite{R11}.  Therefore we concentrate ourselves to the
calculation of formfactors $g_1$ and $f_1$ induced by the magneto-dipole
interaction.

\section{Sum rules for the transition formfactors $f_1(p^2)$ and $g_1(p^2)$}

According to the QCD sum rules ideology, in order to calculate 
the transition 
formfactors $f_1(p^2)$ and $g_1(p^2)$, it is necessary to  write
the representation of a
suitable correlator function in the hadronic and quark-gluon languages.  We
start by considering the following correlator function
\bea
\Pi_\mu(p,q) &=& i \int{d^4 x e^{i p x} \la \gamma(q) \ve \bar q(x)
i \sigma_{\mu \alpha} p_\alpha (1+\gamma_5) b(x) \bar b(0) i \gamma_5 q(0) \ve
0 \ra} 
\eea

This correlator can be calculated in two different ways. On one side we
insert to $\Pi_\mu(p,q)$ the hadronic states with $B$ meson quantum numbers.
Then we have 
\bea
\Pi_\mu(p,q)&=&
\frac{m_{B_q}^2 f_{B_q}}{m_b} \frac{1}{m_{B_q}^2 - (p+q)^2}
\la \gamma(q) \ve \bar q i \sigma_{\mu \alpha} p_\alpha (1+\gamma_5) b
\ve B(p+q) \ra \nnb \\
&=&e \frac{m_{B_q}^2 f_{B_q}}{m_b} \frac{1}{m_{B_q}^2 - (p+q)^2} \times \nnb \\
&&\times \Bigg{\{}
\epsilon_{\mu \alpha \beta \sigma} e_\alpha^*
p_\beta q_\sigma \frac{g_1(p^2)}{m_B^2}+ 
+ i \left[ e_\mu^* (p q) - (e^* p ) q_\mu \right] \frac{f_1(p^2)}{m_B^2} \Bigg{\}}~. 
\eea 
In deriving eq(5) we used 
\bea
\la B \ve \bar b i \gamma_5 q \ve 0 \ra = \frac{m_{B_q}^2 f_{B_q}}{m_b} \nnb
\eea
On the other hand, the correlation function (4), can be calculated in QCD
at large Euclidean momenta $(p+q)^2$.
In general, the correlator (4) can be decomposed into
the parity conserving and parity violating parts
\bea
\Pi_\mu(p,q) = 
\epsilon_{\mu \alpha \beta \lambda} e_\alpha^* p_\beta q_\lambda \Pi_1 + i 
\left[ e_\mu^* (p q) - q_\mu (e^* p) \right] \Pi_2 
\eea 
Equating eqs.(5) and (6) we get sum rules for the formfactors $g_1(p^2)$ and
$f_1(p^2)$.

Let us start calculating  $\Pi_\mu(p,q)$ from QCD side.
The virtuality of the heavy quark in the correlator function under
consideration, is large and of order $m_b^2-(p+q)^2$. Thus, one can use the
perturbative expansion of the heavy quark propagator in the external field
of slowly varying fluctuations inside the photon. The leading contribution is
obtained by using the free heavy quark propagator in eq.(4). Then we have
\bea
\Pi_\mu(p,q)&=&\int{ \frac{d^4x \, d^4k}{(2 \pi)^4} \,
\frac{e^{i(p-k)x}}{(m_b^2-k^2)}}
\la \gamma \ve \bar q(x) i \sigma_{\mu\alpha} p_\alpha
(1+\gamma_5)(\not\!k+m_b)i \gamma_5 q(0) \ve 0 \ra \nnb \\ \nnb \\
&=&- \int{\frac{d^4x \, d^4k}{(2 \pi)^4} \,
\frac{e^{i(p-k)x}}{(m_b^2-k^2)}} \,
p_\alpha \Bigg{\{} m_b \la \gamma \ve\bar q(x) \sigma_{\mu\alpha} (1+\gamma_5)
q(0)\ve 0 \ra - \nnb \\
&&-k_\rho \la \gamma \ve \bar q(x) \sigma_{\mu\alpha} \gamma_\rho (1-\gamma_5)
q(0)
\ve 0 \ra 
\Bigg{\}}
\eea
In this equation a path ordered gauge factor between the quark fields is
omitted, since in the Fock-Schwinger gauge $x_\mu A^\mu(x) = 0$,
where $A^\mu(x)$ is the external electromagnetic field,
it is irrelevant. 

The diagrams (a) and (b) in Fig.2 describe only the short distance
(perturbative) part of these matrix elements corresponding to the photon
emission from the freely  propagating heavy and light quarks.
The non-perturbative contributions
correspond to the propagation of the light quark in the presence of
external electromagnetic field (Fig.2c and 2d).   

We consider now the perturbative contributions. For the diagrams (2-a and
2-b) we can write down the double dispersion representation
\beq
\Pi^{(1,2)} = \int \frac{ds\, dt \, \rho_i^{(1,2)}(s,t)}
{\left[ s-(p+q)^2 \right] (t-p^2)}~ +~\mbox{subtr. terms.}
\eeq
Here, superscripts $1$, and $2$ correspond to
the contributions of the spectral densities to the structures $\epsilon_{\mu
\alpha \beta \lambda} e_\alpha^* p_\beta q_\lambda$ and $e_\mu^* (p q) -
q_\mu (e^* p)$ respectively.

For calculating the spectral
densities $\rho_l$ and $\rho_H$ we use the method given in \cite{R12}. 
After a rigorous calculation for spectral densities, we have
\bea
\rho_l^{(1)}(s,t)&=& -\frac{N_c}{16 \pi^2} e e_q \, s \delta(t-s) 
\ga 1 - \frac{m_b^4}{s^2} \dr~, \\
\nnb \\
\rho_H^{(1)}(s,t)&=& - \frac{2 N_c}{16 \pi^2}  e e_b \,
s \delta(t-s)\ga 1 - \frac{m_b^2}{s} \dr~, \\ \nnb \\
\rho_l^{(2)}(s,t)&=&\frac{2 N_c}{16 \pi^2} e e_q \Bigg{\{} \delta(t-s)
\ga 1- \frac{m_b^2}{s} \dr
\ga - \frac{m_b^2}{2} + \frac{3}{2} s \dr  - \nnb \\ \nnb \\
&& - \delta'(t-s) \ga 1- \frac{m_b^2}{s} \dr \ga s - 
m_b^2 \dr s \Bigg{\}}~, \\ \nnb \\
\rho_H^{(2)}(s,t)&=&\frac{2 N_c}{16 \pi^2}  e e_b \Bigg{\{} \delta(t-s)
\left[ \ga 1- \frac{m_b^2}{s} \dr
\ga \frac{3}{2} s + \frac{1}{2} m_b^2 \dr - 
2 m_b^2 \, ln \ga \frac{s}{m_b^2} \dr \right] - \nnb \\ \nnb \\
&& - \delta'(t-s) \left[ \ga 1- \frac{m_b^2}{s} \dr \ga s^2 +  s m_b^2 \dr -
2 s m_b^2 \, ln \ga \frac{s}{m_b^2} \dr \right] \Bigg{\}}~.
\eea
In eqs.(9-12) $\rho_l$ and $\rho_H$ corresponds to the interaction of the
photon with the light and $b$ quarks, 
$N_c = 3$ is the color factor, $e_q$ and $e_b$ the electric charge of
the light and $b$ quarks and $m_b$ is the mass of the $b$-quark
$\delta'(t-s) = \frac{d}{dt} \delta(t-s)$.

Next consider the non-perturbative contributions.
From eq.(7) it follows that the non-perturbative contributions are expressed
via the matrix elements of the gauge invariant nonlocal operators,
sandwiched in
between the vacuum and the photon state. These matrix elements define the
following light cone photon wave functions (\cite{R10,R13}, see also the
first reference in  \cite{R11}): 
\newpage
\bea
\la \gamma \ve \bar q(x) \sigma_{\mu \alpha} q(0) \ve 0 \ra &=& i e e_q \la \bar
q q \ra \int_0^1 du e^{i u q x} \nnb \\
&&\Big{\{} (e_\mu q_\alpha - e_\alpha q_\mu)
\left[ \chi \phi ( u ) + x^2 ( g_1(u) - g_2(u)) \right] + \nnb \\
&& + g_2(u) \left[ q x (e_\mu x_\alpha - e_\alpha x_\mu) + 
e x (x_\mu q_\alpha - x_\alpha q_\mu) \right] \Big{\}}
~~~~~\mbox{and} \nnb \\
\la \gamma \ve \bar q(x) \gamma_\mu \gamma_5 q(0) \ve 0 \ra &=& \frac{1}{4}
e \epsilon_{\mu \alpha \beta \lambda} e_\alpha p_\beta x_\lambda f \int_0^1
g_\perp (u) e^{i u q x}~.
\eea
Here $\chi$ is magnetic susceptibility of the quark condensate, $\phi (u)$ ,
$g_\perp (u)$ are the leading twist $\tau = 2$ photon wave functions, $g_1(u)$ and
$g_2(u)$ are the two particle $\tau = 4$ wave functions.  Note that for
calculating the matrix elements
\bea
&&
\la \gamma (q) \ve \bar q(x) \sigma_{\mu \alpha} \gamma_\rho (1-\gamma_5) q(0) \ve
0 \ra \nnb~~~\mbox{and} \nnb \\
&&\la \gamma (q) \ve \bar q(x) \sigma_{\mu \alpha} \gamma_5 q(0) \ve 0 \ra \nnb
\eea 
we use the following identities:
\bea
&&\sigma_{\mu \alpha} \gamma_5 = \frac{1}{2} i \epsilon_{\mu \alpha \lambda
\rho} \sigma_{\lambda \rho} ~,\\
&&\sigma_{\mu \alpha} \gamma_\rho = i (\gamma_\mu g_{\alpha \rho} -
\gamma_\alpha g_{\mu \rho}) + \epsilon_{\mu \alpha \rho \lambda}
\gamma_\lambda \gamma_5~.
\eea

After lengthy calculations for $\Pi_1$ and $\Pi_2$ we get the following
results, which describe the non-perturbative contributions:

\bea
\Pi_1&=& m_b e e_q \la \bar q q \ra \int_{0}^{1} du \Bigg{\{} - \frac{\chi \phi
(u)}{\Delta} + 8 m_b^2 \frac{g_1(u) - g_2(u)}{\Delta^3} - \frac{4(m_b^2 - p^2)}
{\Delta^3} g_2 \Bigg{\}} - \nnb \\ \nnb \\
&& - \frac{e}{4} f \int_{0}^{1} du \Bigg{[} \frac{1}{\Delta} + \frac{p^2 +
m_b^2}{\Delta^2} \Bigg{]} g_\perp(u) + e e_b m_b \la \bar q q \ra
\frac{1}{(m_b^2 - p^2) \left[ m_b^2 - (p+q)^2 \right]} ~,\\ \nnb \\
\Pi_2&=& m_b e e_q \la \bar q q \ra \int_{0}^{1} du \Bigg{\{} - \frac{\chi \phi(u)}
{\Delta} + 8 m_b^2 \frac{g_1(u) - g_2(u)}{\Delta^3} - \frac{4(m_b^2 + p^2)}
{\Delta^3} g_2 \Bigg{\}} - \nnb \\ \nnb \\
&& - \frac{e}{4} f \int_{0}^{1} du \Bigg{[} \frac{2}{\Delta} + \frac{2 u p q}
{\Delta^3} \Bigg{]} g_\perp(u) + e e_b m_b \la \bar q q \ra
\frac{1}{(m_b^2 - p^2) \left[ m_b^2 - (p+q)^2 \right]}
\eea

Here $\Delta = m_b^2 - ( p + u q)^2$. Last term in eqs.(16) and (17) describes
the case when a photon is emitted from the heavy quark (see Fig.2d).
Collecting eqs.(9-12) and (16-17) we finally get the following expressions
 for the invariant functions $\Pi_1$ and $\Pi_2$:

\bea
\Pi_1&=& - \frac{N_c e}{16 \pi^2} \int_{0}^{1} du 
\frac{m_b^2-p^2 \bar u}{u^2 \Delta_1} 
\ga 1 - \frac{m_b^2 u}{m_b^2 - p^2 \bar u} \dr
\left[ e_q \ga 1+ \frac{m_b^2 u}{m_b^2 - p^2 \bar u} \dr
+ 2 e_b \right] + \nnb \\ \nnb \\
&+& m_b e e_q \la \bar q q \ra \int \frac{du}{u} \Bigg{\{} - \frac{\chi
\phi(u)}{\Delta_1} + 8 m_b^2 \frac{g_1(u) - g_2(u)}{u^2 \Delta_1^3} - 4 \frac{(m_b^2
- p^2)}{u^2 \Delta_1^3} g_2 \Bigg{\}} - \nnb \\ \nnb \\
&-& \frac{e}{4} f \int \frac{du}{u} \Bigg{[} \frac{1}{\Delta_1} + \frac{p^2
+m_b^2}{u \Delta_1^2} \Bigg{]} g_\perp(u) + e e_b m_b \la \bar q q \ra
\frac{1}{(m_b^2 - p^2) \left[ m_b^2 - (p+q)^2 \right]}~,  \\ \nnb \\ \nnb \\
\Pi_2&=&
\frac{N_c e}{16 \pi^2} 
\int_{0}^{1} \frac{du}{\Delta_1 \ga m_b^2-p^2 \dr} 
\Bigg{\{} \ga 1 - \frac{u m_b^2}{m_b^2-p^2 \bar u} \dr  \Bigg{[} (e_b +
e_q) \frac{m_b^2-p^2 \bar u}{u} \ga \frac{m_b^2-p^2 \bar u}{u} - 3 p^2 \dr + \nnb \\
\nnb \\
&+& (e_q - e_b) m_b^2 \ga \frac{m_b^2-p^2 \bar u}{u} + p^2 \dr \Bigg{]} 
+ 4 e_b m_b^2 p^2 ln \ga \frac{m_b^2-p^2 \bar u}{m_b^2 u} \dr \Bigg{\}}
+ \nnb \\ \nnb \\
&+& m_b e e_q \la \bar q q \ra \int_{0}^{1} \frac{du}{u} \Bigg{\{} - \frac{\chi
\phi(u)}{\Delta_1} + 8 m_b^2 \frac{g_1(u) - g_2(u)}{u^2 \Delta_1^3} + 4
\frac{\ga m_b^2
+ p^2 \dr}{u^2 \Delta_1^3} g_2 \Bigg{\}} - \nnb \\ \nnb \\
&-& \frac{e}{4} f \int_{0}^{1} \frac{du}{u} \Bigg{[} \frac{2}{\Delta_1} +
\frac{2 (p q)}{u \Delta_1^3}\Bigg{]} g_\perp(u) + e e_b m_b \la \bar q q
\ra\frac{1}{\ga m_b^2 - p^2 \dr \left[ m_b^2 - (p+q)^2 \right]}~, \\ \nnb
\eea  
where $\Delta_1 = \frac{(m_b^2 - p^2 \bar u)}{u}-(p+q)^2$, $\bar u = 1-u$.
In eqs.(18) and (19) we have rewritten the dispersion integral in terms of
the variable $u = (m_b^2 - p^2) / (s - p^2)$ .

Here we would like to make the following remark. As we noted earlier, 
the functions $g_1(u)$ and $g_2(u)$ represent twist $\tau=4$ contributions
to the two-particle photon wave function. To this accuracy, in eq.(19)
we must take into account other twist $\tau = 4$ photon wave functions (see for
example \cite{R17}). Using the equation of motion, one
can relate them to the three-particle wave functions of twist $\tau=4$ with
an additional gluon from heavy quark \cite{R17}. But, these 
three-particle wave function contributions, in general, are small and we will
neglect them in further analysis.

The remaining  task is now to match eqs.(18) and (19)
with the corresponding hadronic
representation (see eq.(5))
and to extract the formfactors $g_1(p^2)$ and $f_1(p^2)$.
As usual, invoking duality, we assume that above certain threshold
$s_0=35~GeV^2$ (this value follows from two-point sum rules analysis) the
spectral density $\rho(s)$ associated with higher resonances and continuum
states coincides with the spectral density from perturbative part. This
procedure is equivalent to writing $(m_b^2 - p^2)/(s_0 - p^2)$ in the lower
limit of the  integration over $u$ in eqs.(18) and (19) (for more detail
see \cite{R11,R15}). Finally applying the Borel
transformation on the variable  $-(p+q)^2 \rar M^2$ to suppress both higher state
resonances and higher Fock states in the full photon wave functions, we get the
following sum rules for the
formfactors:
\bea
g_1 \ga p^2 \dr &=&
- \frac{m_b}{f_B}~ e^{\frac{m_B^2}{M^2}}~ \Bigg{\{}
\frac{N_c}{16 \pi^2} \int_\delta^{1} \frac{du}{u^2}
\ga m_b^2-p^2 \bar u \dr \ga 1 - \frac{m_b^2 u}{m_b^2 - p^2 \bar u} \dr 
\times \nnb \\ 
&&\times \left[ e_q \ga 1+ \frac{m_b^2 u }{m_b^2 - p^2 \bar u} \dr + 2 e_b \right]
e^{-\frac{\ga m_b^2-p^2 \bar u \dr}{u M^2}}  + \nnb \\
&+& m_b \la \bar q q \ra e_q \int_\delta^1  \frac{du}{u} 
\left[ \chi \phi (u) - 4 m_b^2 \ga g_1 - g_2 \dr \frac{1}{u^2 M^4}
-2 \frac{ \ga m_b^2 - p^2 \dr}{u^2 M^4} g_2
\right] e^{- \frac{\ga m_b^2-p^2 \bar u \dr}{u M^2} } + \nnb \\
&+& \frac{f}{4} \int_{\delta}^{1} du \frac{g_\perp (u)}{u}
\ga 1 + \frac{p^2+m_b^2}{u M^2} \dr 
e^{-\frac{ \ga m_b^2 - p^2 \bar u \dr}{u M^2}} - e_b m_b
\frac{\la \bar q q \ra}{m_b^2-p^2} e^{- \frac{m_b^2}{M^2}}\Bigg{\}}~, \nnb \\
\nnb \\ \nnb \\
f_1\ga p^2 \dr &=&
\frac{m_b}{f_B}~ e^{\frac{m_B^2}{M^2}}~ \Bigg{\{}
\frac{N_c}{16 \pi^2} \int_\delta^1 \frac{du}{m_b^2-p^2}
e^{- \frac{\ga m_b^2-p^2 \bar u \dr}{u M^2} } 
\Bigg{[}
\ga 1 - \frac{m_b^2 u}{m_b^2 - p^2 \bar u} \dr \times \nnb \\ \nnb \\
&&\times \ga \ga e_q+e_b \dr
\frac{m_b^2 - p^2 \bar u}{u} \ga \frac{m_b^2 - p^2 \bar u}{u}-3 p^2 \dr
 +\ga e_q-e_b \dr m_b^2 \ga \frac{m_b^2 - p^2 \bar u}{u}
+p^2 \dr \dr \nnb \\ \nnb \\
&+&4 m_b^2 p^2 ln \, \frac{m_b^2 - p^2 \bar u}{m_b^2 u} 
\Bigg{]} + m_b e_q \la \bar q q \ra \int_{\delta}^{1} \frac{du}{u}
\Bigg{[} - \chi \phi (u) + \frac{4 m_b^2}{M^4 u^2} \ga g_1 - g_2 \dr
+\nnb \\ \nnb \\
&+&  \frac{2 \ga p^2+m_b^2 \dr}{u^2 M^4 } g_2 \Bigg{]} 
e^{- \frac{\ga m_b^2-p^2 \bar u \dr}{u M^2} }+ \nnb \\
&+&\frac{f}{4} \int_{\delta}^{1} \frac{du}{u} g_\perp (u)
\ga  -1 +\frac{p^2-m_b^2 }{u M^2 }\dr 
e^{- \frac{\ga m_b^2-p^2 \bar u \dr}{u M^2} } +e_b m_b
 \frac{\la \bar q q \ra}{m_b^2-p^2}e^{- \frac{m_b^2}{M^2}}
\Bigg{\}}
\eea

At the end of this section we give  the result for the differential
decay widths:

\bea
\frac{d \Gamma}{d \hat{s}} &=& \frac{\alpha^3 G^2}{768 \pi^5} \left| V_{tb}
V_{tq}^* \right|^2 m_B^5 \hat{s} ( 1-\hat{s})^3 
\sqrt{1 - 4 \frac{m_l^2}{m_B^2 \hat{s}}} \times \nnb \\
&&\times \Bigg{\{} \frac{1}{m_B^2} \left[ \left| A \right|^2 + \left| B \right|^2 \right]
+\frac{1}{m_B^2} \left| C_{10} \right|^2 \left[ f^2(p^2) + g^2(p^2) \right]
\Bigg{\}}~,
\eea
\newpage
where
\bea
\hat{s} &=& p^2/m_B^2~, \nnb \\
A &=& C_9^{eff} g(p^2) - 2 \, C_7 \frac{m_b}{p^2} g_1(p^2)~,~\mbox{and} \nnb \\
B &=& C_9^{eff} f(p^2) - 2 \, C_7 \frac{m_b}{p^2} f_1(p^2)~. \nnb
\eea

\section{Numerical Analysis}

For calculating formfactors $f_1(p^2)$ and $g_1(p^2)$ we use the following input
parameters: \\
$m_b = 4.7~GeV,~s_0 \simeq 35~ GeV^2,~ f_B = 140~ MeV
~~\mbox{\cite{R14,R15}} ,~\phi(u) = 6 u (1-u)$ 
\cite{R16,R17}. To the leading twist accuracy we
use for $g_\perp (u) = 1$ (see first reference in \cite{R11} ) and for $g_1(u)$ and $g_2 (u)$
the following expressions \cite{R13}:
\bea
g_1(u) &=& - \frac{1}{8} (1-u)(3-u) \\
g_2(u) &=& - \frac{1}{4} (1-u)^2
\eea

The magnetic susceptibility $\chi$ was determined in \cite{R18}, $\chi =
- 3.4~ \mbox{GeV} ^{-2}$ at the scale $ \mu_b \sim \sqrt{m_B^2 - m_b^2} ~~,~~
\la \bar q q \ra = - (0.26 ~\mbox{GeV} ~)^3$ . The Borel parameter $M^2$ has
been varied in the region $ 8 ~\mbox{GeV}^2 < M^2 < 20  ~\mbox{GeV}^2$ .
Numerical analysis shows that the variation of $M^2$ in this region,
changes the results by less than $8 \% $ . The predictions of the sum rules
are very stable in this region of the Borel parameter and vary only a few
percent with the changes of $m_b,~ s_0$ and $f_B$ within the intervals
allowed by the two point sum rules for $f_B$.

The sum rules is reliable in the region $m_b^2 - p^2 \sim$ a few GeV$^2$,
which is smaller than $p^2 = m_b^2$. In order to extent our results to the
whole region of $p^2$ we use some extrapolation formulas. We found that the
best agreement is achieved by the dipole type formulas

\bea
g_1(p^2) &=& \frac{3.74~GeV^2}{(1-\frac{p^2}{m_1^2})^2}~, \\ \nnb  \\
f_1(p^2) &=& \frac{0.68~GeV^2}{(1-\frac{p^2}{m_2^2})^2}~,  
\eea
where $m_1^2 = 40.5~GeV^2$ and $m_2^2 = 30~ GeV^2$.
For calculating differential and total decay widths, we need the values of
$C_9^{eff}~,~ C_7$ and $C_{10}$ coefficients and the explicit forms of the
formfactors $g(p^2)$ and $f(p^2)$. These formfactors are calculated in \cite{R11}: 
\bea
g_(p^2) &=& \frac{1~GeV}{(1-\frac{p^2}{5.6^2})^2}~, \\ \nnb  \\
f_(p^2) &=& \frac{0.8~GeV}{(1-\frac{p^2}{6.5^2})^2}~.  
\eea
The values of the Wilson
coefficients $C_7$ and $C_{10}$ are taken from \cite{R9,R10} as
\bea
C_7 = -0.315~~,~~C_{10} = - 4.642~, \nnb
\eea 
and the expression $C_9^{eff}$ for $b \rar s$ transition, in 
the next-to-leading order approximation is given as 
(see \cite{R19})
\bea
C_9^{eff} &=& C_9 + 0.124 w(\hat{s}) + g(\hat{m_c},\hat{s})( 3 C_1 + C_2 + 3
C_3 + C_4 + 3 C_5 + C_6) - \nnb \\
&& - \frac{1}{2} g(\hat{m_q},\hat{s})(C_3 + 3 C_4)
-\frac{1}{2} g(\hat{m_b},\hat{s})(4 C_3 + 4 C_4 + 3 C_5 + C_6) + \nnb \\
&& + \frac{2}{9} (3 C_3 + C_4 + 3 C_5 + C_6)~,
\eea
with
\bea
C_1 &=& -0.249 ~~,~~ C_2 = 1.108 ~~,~~ C_3 = 1.112 \times 10^{-2} ~~,~~ C_4 =
-2.569 \times 10^{-2} \nnb \\
&&C_5 = 7.4 \times 10^{-3} ~~,~~ C_6 =-3.144 \times 10^{-2}~~,~~C_9 = 4.227. \nnb
\eea
The value of $C_9^{eff}$ for $b \rar d$ transition, can be obtained by
adding to eq.(28) the term $\lambda_u \left[ g(\hat{m_c},\hat{s})-
g(\hat{m_d},\hat{s}) \right] \ga 3 \, C_1 + C_2 \dr$, where
\bea
\lambda_u = \frac{V_{ub} V^*_{ud}}{V_{tb} V^*_{td}}~. \nnb
\eea
For obtaining these values we used $\Lambda_{QCD} = 225 MeV,~ sin^2\theta_W
=0.23,~ m_t = 176~GeV ,~ m_W = 80.2~ GeV$ and $\hat{m_q} = m_q / m_b$. In
the above formula $w(\hat{s})$ represents the one-gluon correction to the
matrix element $O_9$ and explicit expression can be found in \cite{R10},
while the function $g(\hat{m_q},\hat{s})$ arises from the one loop
contributions of the four quark operators $O_1$ -- $O_6$ (see for example
\cite{R9,R10}), i.e.

\bea
g(\hat{m_q},\hat{s'})&=& - \frac{8}{9} ln \hat{m_q} + \frac{8}{27} +
\frac{4}{9} y_q - \frac{2}{9} (2 + y_q) \sqrt{11-y_q} +
+\Bigg{\{} \Theta(1-y_q) \times \nnb \\ \nnb \\
&& \times \ga ln \frac{1+\sqrt{1-y_q}}{1-\sqrt{1-y_q}} - i
\pi \dr + \Theta(y_q-1) arctg \frac{1}{\sqrt{y_q-1}} \Bigg{\}}
\eea
with $y_q = \hat{m_q}^2 / \hat{s'}$, and $\hat{s'}=p^2/m_b^2$.

For a more complete analysis of the above decay, one has to take into account the
long distance contributions. In the case of the $J/\psi$ family, this is
accomplished by introducing a Breit-Wigner formula through the replacement
(see \cite{R20})
\bea
g(\hat{m_c},\hat{s'}) \rar g(\hat{m_c},\hat{s'}) - \frac{3 \pi}{\alpha^2}
\sum_{V=J/\psi , \psi'} \frac{\hat{m_V} Br( V \rar l^+ l^-)
\hat{\Gamma}_{tot}^V}{\hat{s'} - \hat{m_V}^2 + i \hat{m_V}
\hat{\Gamma}_{tot}^V}
\eea
where $\hat{m_V} = m_V/m_b$, $\hat{\Gamma}_{tot} = \Gamma /m_b$ . The masses
and decay widths of the corresponding mesons are listed in \cite{R21}.
In Fig.3 we present the differential decay rate for $B_s \rar \mu^+ \mu^-
\gamma$ decay (behavior of the differential decay rate for $B_s \rar e^+
e^-\gamma$ decay is similar) as a function of
$\hat{s}$, with and without resonance ($J/\psi~ and~ \psi'$)
contributions. From this figure we see that
the contribution from soft photons, corresponding to large $\hat{s}$ region is
negligible.

Using the above mentioned values of the parameters and $\ve V_{tb}
V_{ts}^*\ve = 0.045$ , $\ve V_{tb} V_{td}^*
\ve = 0.01$, $\tau(B_s) = 1.34 \times 10^{-12}~s$ , 
$\tau(B_d) = 1.5 \times 10^{-12}~s$
\cite{R21}, for branching ratios we get (without the long distance contributions):

\bea
B(B_s \rar e^+ e^- \gamma) &=& 2.35 \times 10^{-9} \nnb \\
B(B_s \rar \mu^+ \mu^- \gamma) &=& 1.9 \times 10^{-9} \nnb \\
B(B_d \rar e^+ e^- \gamma) &=& 1.5 \times 10^{-10} \nnb \\
B(B_d \rar \mu^+ \mu^- \gamma) &=& 1.2 \times 10^{-10} 
\eea
For comparison we present also the constituent model prediction (at $f_B =
140~MeV$ , $m_s = 0.57~GeV$ , $m_d = 0.35~GeV$) \cite{R7}:
\bea
B(B_s \rar e^+ e^- \gamma) &=& 3 \times 10^{-9} \nnb \\
B(B_s \rar \mu^+ \mu^- \gamma) &=& 2.3 \times 10^{-9} \nnb \\
B(B_d \rar e^+ e^- \gamma) &=& 4 \times 10^{-10} \nnb \\
B(B_d \rar \mu^+ \mu^- \gamma) &=& 3 \times 10^{-10} 
\eea

We see that the constituent quark model and light cone sum rules method
predictions on the branching ratios are very close.
Let us compare our results on branching ratios with those of pure leptonic
decays.
The rates for the pure leptonic decays are (see for example
\cite{R6,R7})
\beq
\Gamma ( B_q \rar l^+ l^-) =
\frac{\alpha^2 G_F^2 f_{B_q}^2 m_{B_q} m_l^2}{16 \pi^3}
\ve V_{tb} V_{tq}^* \ve^2 C_{10}^2 \nnb
\eeq
If we use the value of $f_{B_s} \simeq f_{B_d} \simeq 140~MeV$, for the
corresponding Branching ratios we get:
\bea
B(B_s \rar e^+ e^-) &=& 3 \times 10^{-14} \nnb \\
B(B_s \rar \mu^+ \mu^-) &=& 1.3 \times 10^{-9} \nnb \\
B(B_d \rar e^+ e^-) &=& 2.1 \times 10^{-15} \nnb \\
B(B_d \rar \mu^+ \mu^-) &=& 9 \times 10^{-11}
\eea
From these values and eq.(30) it follows that, the radiative decays dominate over
the pure leptonic decays in the corresponding channels and $B_s \rar e^+ e^-
\gamma$ decay mode has a larger branching ratio.
Few words about the experimental detectabilty of these processes is in
order. In future $B$-factories and LHC approximately
$6 \times 10^{11}(2 \times 10^{11})~B_d(B_s)$ mesons are expected per   
year. Therefore the decays $B_{s(d)} \rar l^+ l^- \gamma$ are expected to be  
quite detectable in these machines.

In conclusion, we have analyzed the rare $B_q \rar l^+ l^- \gamma$ decays in SM
and obtain the branching ratios  for $B_s \rar l^+ l^- \gamma$
to be around $2 \times 10^{-9}$ and around $ 2 \times 10^{-10}$ for
$B_d \rar l^+ l^- \gamma$.

\section*{Acknowledgements}
We thank Prof. M. P. Rekalo for helpful discussions.

\newpage

\section*{Figure Captions}
{\bf 1.} Feynman diagrams in the Standard Model for $b \bar q \rar l^+ l^-$
\\
{\bf 2.} Diagrams describing the perturbative and non-perturbative 
contributions  to the correlator function (4).\\
{\bf 3.} Differential decay rates of $B_s  \rar \mu^+ \mu^- \gamma$ versus 
$\hat{s}=p^2/m_B^2$. Here the thick line corresponds to the case without the 
$J/\psi$, $\psi'$ and the thin line with the $J/\psi$, $\psi'$
contributions.


\newpage 

\begin{thebibliography}{99}

\bibitem{R1} A. Ali, {\it Prep.} {\bf DESY 96-106} (1996),
to appear in the {\bf proc. XXX.} Nathiagali Summer
College on Physics and Contemporary Needs, Nova
Science Publ., NY, Editors: Riazuddin, K. A. Shoaib {\it et. al.}; \\
A. J. Buras, M. K. Harlander, {\bf Heavy Flavors} p. 58-201
Editors: A. J. Buras, \\
M. Lindner, (World Scientific, Singapore);
A. Ali, {\it Nucl. Phys.} {\bf B}, {\it Proc. Supp.} {\bf 39 BC} (1995)
408-425; S. Playfer and S. Stone, {\it Int. J. Mod. Phys.} {\bf A10} (1995)
4107.

\bibitem{R2} Z. Ligeti and M. Wise, 
{\it Phys. Rev.} {\bf D53} (1996) 4937.

\bibitem{R3} R. Ammar {\it et. al.} 
{\bf CLEO} Collaboration, {\it Phys. Rev. Lett.} {\bf 71} (1993) 674.

\bibitem{R4} M. S. Alam {\it et. al.}, 
{\bf CLEO} Collaboration, {\it Phys. Rev.Lett.} {\bf 74} (1995) 2885.

\bibitem{R5} B. A. Campbell and P. J. O'Donnell,
{\it Phys. Rev.} {\bf D25} (1982) 1989; \\
A. Ali, in {\bf B decays}, Editor: S. Stone (World Scientific, Singapore)
67.

\bibitem{R6} G. Buchalla and A. J. Buras,
{\it Nucl. Phys.} {\bf B400} (1993) 225.

\bibitem{R7} G. Eilam, Cai-Dian L\"{u}, and  Da-Xin-Zhang, {\it Prep.}
{\bf Technion}-PH-96-12, \\ 
{\bf hep-ph}/9606444 (1996).

\bibitem{R8} V. M. Braun, {\it Prep.} {\bf NORDITA}-95-69-P(1995),
{\bf hep-ph}/9510404 (1995), to appear in: {\it Proc. of the Int. Europhys.}
Conf. on High Energy Physics, Brussels, Belgium, 1995.

\bibitem{R9} G. Buchalla, A. J. Buras and M. E. Lautenbacher,
{\it Prep.}
{\bf MPI-ph}/95-104, \\
{\bf TUM}-T31-100/95, {\bf Fermilab-PUB}-95/305-T,
{\bf SLAC-PUB} 7009, \\
{\bf hep-ph}/9512380 (1995);
A. J. Buras and M. M\"{u}nz, 
{\it Phys. Rev.} {\bf D52} (1995) 186.

\bibitem{R10} M. Misiak, 
{\it Nucl. Phys.} {\bf B398} (1993) 23;   
Erratum: {\it ibid} {\bf B439} (1995) 461.

\bibitem{R11} G. Eilam, I. Halperin and R. R. Mendel
{\it Phys. Lett.} {\bf B361} (1995) 137. \\
T. M. Aliev, A. \"{O}zpineci and M. Savc{\i},
{\it Phys. Lett.} {\bf B} (in press).

\bibitem{R12} V. A. Beylin and A. V. Radyushkin,
{\it Nucl. Phys.} {\bf B260} (1985) 61; \\
P. Ball and V. M. Braun, 
{\it Phys. Rev.} {\bf D49} (1994) 2472.

\bibitem{R13} A. Ali and V. M. Braun,
{\it Phys. Lett.} {\bf B359} (1995) 223; \\
A. Khodjamirian, G. Stoll and D. Wyler,    
{\it Phys. Lett.} {\bf B358} (1995) 129.   

\bibitem{R14} T. M. Aliev and V. L. Eletsky,
{\it Sov. Nucl. Phys.} {\bf 38} (1983) 936. 

\bibitem{R15} V. M. Belyaev, V. M. Braun,
A. Khodjamirian and R. R\"{u}ckl, \\
{\it Phys. Rev.} {\bf D51} (1995) 6177.

\bibitem{R16} V. L. Chernyak and I. R. Zhitnitsky,
{\it Phys. Rep.} {\bf C112} (1984) 173; \\
A. V. Efremov and A. V. Radyushkin, 
{\it Phys. Lett.} {\bf B94} (1980) 245; \\
G. P. Lepage and S. J. Brodsky, 
{\it Phys. Lett.} {\bf B87} (1979) 359.

\bibitem{R17} I. I. Balitsky, V. M. Braun and A. V. Kolesnichenko,
{\it Nucl. Phys.} {\bf B312} (1989) 509. \\
V. M. Braun and I. B. Filyanov, {\it Z. Phys.} {\bf C44} (1989) 157.

\bibitem{R18} V. M. Belyaev and  Y. I. Kogan, {\it Sov. Nucl. Phys.}
{\bf 40} (1984) 659; \\
I. I. Balitsky, A. V. Kolesnichenko and A. Y. Yung,
{\it Sov. Nucl. Phys.} {\bf 41} (1985) 178.
 
\bibitem{R19} F. Kr\"{u}ger and L. M. Sehgal,
{\it Phys. Lett.} {\bf B380} (1996) 199.

\bibitem{R20} C. S. Lim, T. Morozumi and A. I. Sanda,
{\it Phys. Lett.} {\bf B218} (1989) 343; \\
N. G. Deshpande, J. Trampetic and K. Panose,
{\it Phys. Rev.} {\bf D39} (1989) 1461; \\  
P. J. O'Donnell and H. K. K. Tung, 
{\it ibid} {\bf D43} (1991) R2067; \\
P. J. O'Donnell, M. Sutherland and H. K. K. Tung,
{\it ibid} {\bf D46} (1992) 4091;\\
A. I. Vainshtein, V. I. Zakharov, L. B. Okun and M. A. Shifman,\\
{\it Sov. J. Nucl. Phys.} {\bf 24} (1976) 427.

\bibitem{R21} Particle Data Group, R. M. Barnett {\it et. al.},
{\it Phys. Rev.} {\bf D54} (1996) 1.

\end{thebibliography}
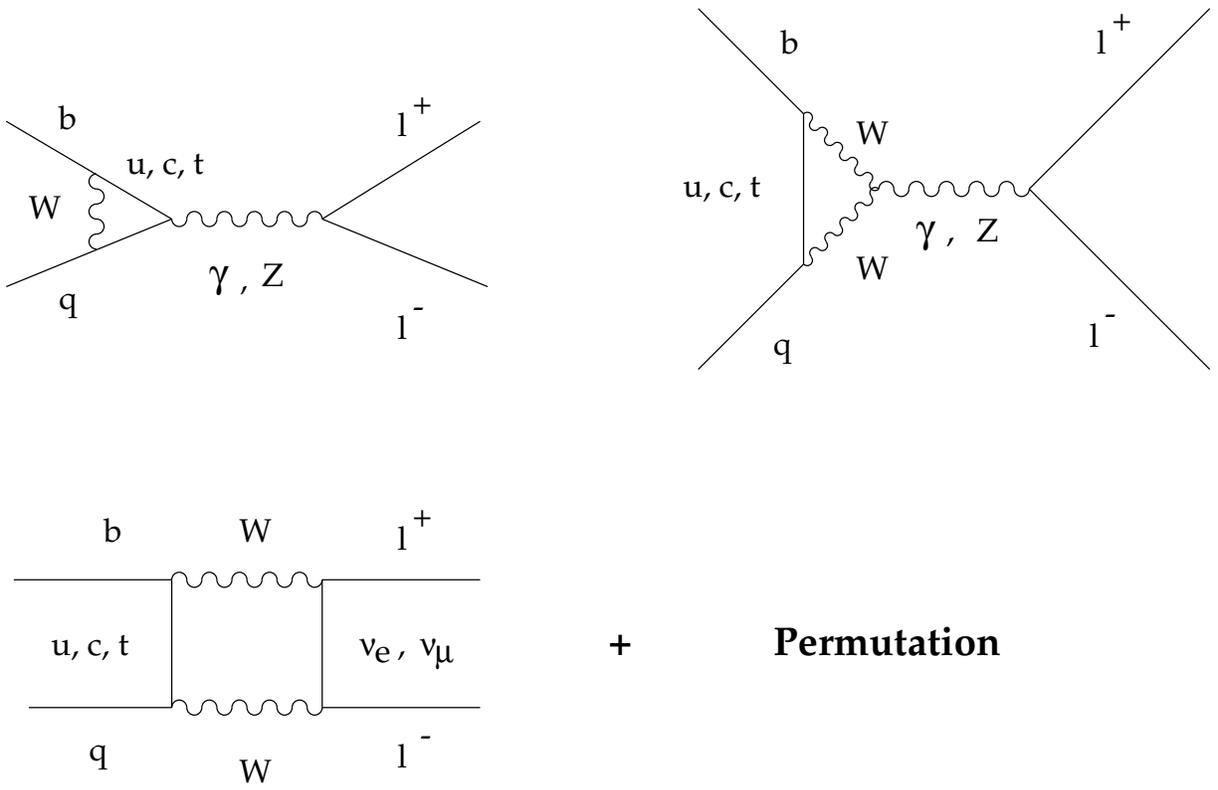
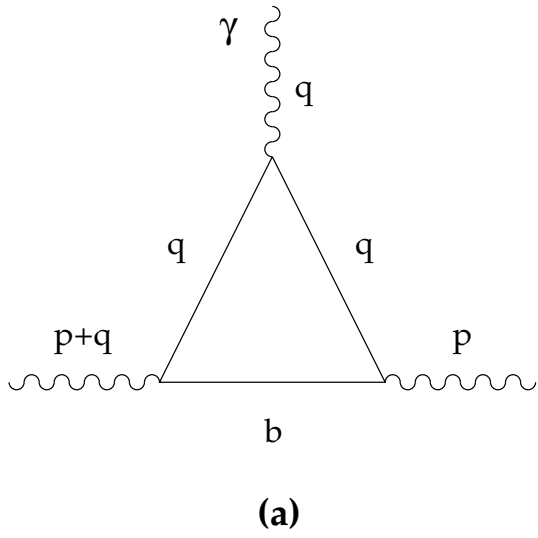
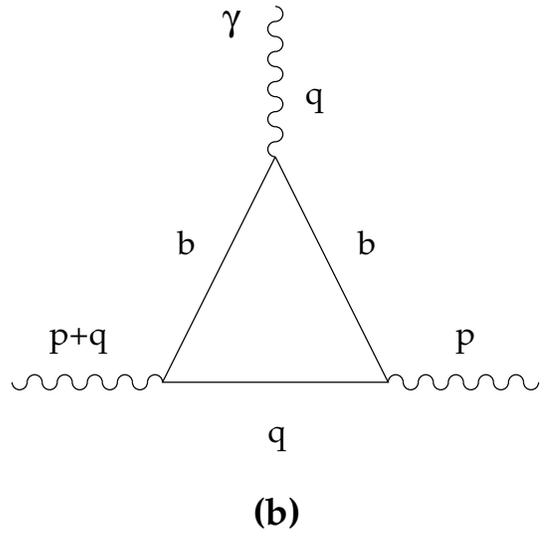
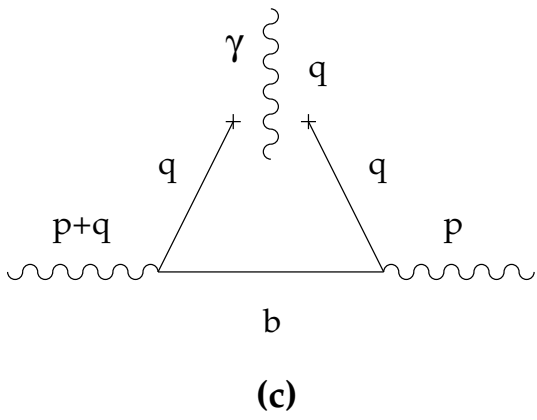
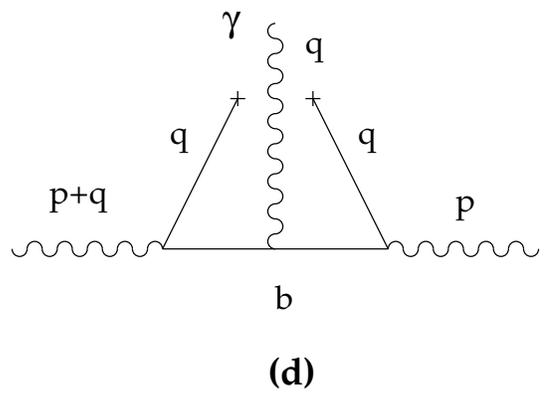
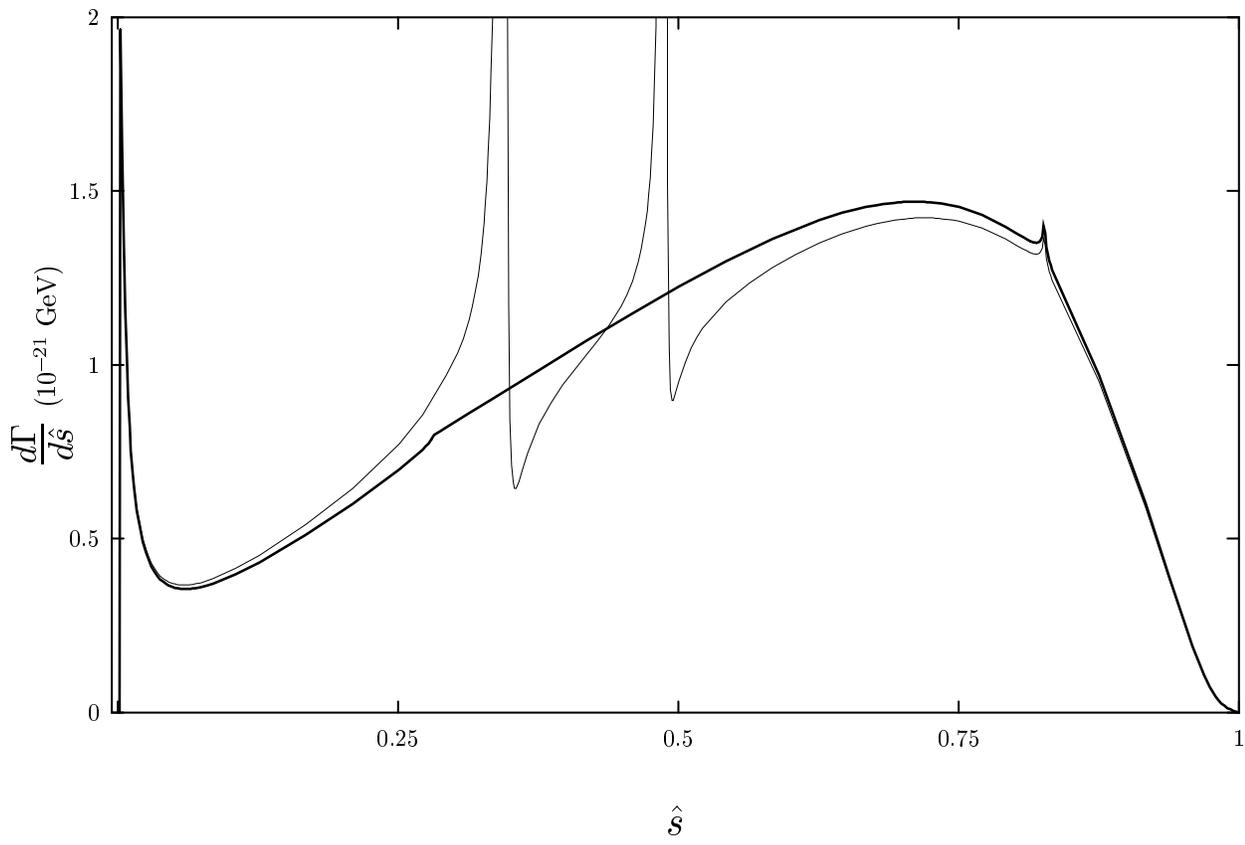
\end{document}